\documentstyle[12pt]{article}
\newcommand{\be}{\begin{equation}}
\newcommand{\ee}{\end{equation}}
\begin{document}
\begin{center}
{\bf Black Hole Formation in Bidimensional Dilaton Gravity\\
 Coupled to Scalar Matter Systems}\footnote{This work is supported in part
 by funds provided by the U.S. Department of Energy (D.O.E.) under
 cooperative research agreement DE-FC02-94ER40818, by Conselho
 Nacional de Desenvolvimento Cient\'\i fico e Tecnol\'ogico, CNPq, and
Funda\c c\~ao Universit\'aria Jos\'e Bonif\'acio, FUJB, Brazil.}
\end{center}
\begin{center}
{M. Alves$^a$, D. Bazeia$^{b,c}$ and V. B. Bezerra$^c$}
\end{center}
\begin{center}
{$^a$Instituto de F\'\i sica, Universidade Federal do Rio de Janeiro\\
Caixa Postal 68528, 21945-970 Rio de Janeiro, Brazil}
\end{center}
\begin{center}
$^b$Center for Theoretical Physics\\
Laboratory for Nuclear Science and Department of Physics\\
Massachusetts Institute of Technology, Cambridge, Massachusetts 02139-4307
\end{center}
\begin{center}
$^c$Departamento de F\'\i sica, Universidade Federal da Para\'\i ba\\
Caixa Postal 5008, 58051-970 Jo\~ao Pessoa, Para\'\i ba, Brazil
\end{center}

\vskip 1cm

\begin{center}
Abstract
\end{center}

This work deals with the formation of black hole in bidimensional dilaton
gravity coupled to scalar matter fields. We investigate two scalar matter
systems, one described by a sixth power potential and the other defined with
two scalar fields containing up to the fourth power in the fields. The
topological solutions that appear in these cases allow the formation of
black holes in the corresponding dilaton gravity models.
\vskip 2cm
\begin{center}
\end{center}

\newpage

\section{Introduction}

The coupling of scalar matter fields with bidimensional
dilaton gravity, originally proposed by Callan, Giddings, Harvey and
Strominger   (CGHS) \cite{cgh92} has attracted attention due to connections
with black hole physics \cite{bol91}, specially
within the context of formation and evaporation of black holes. In cosmology
the dilaton field appears to be important, and so several mechanisms for
cosmological dilaton production has been discussed in the literature
\cite{rth95}. The CGHS model has its origin in the dimensional reduction of
the more realistic, four dimensional, Einstein-Hilbert gravitation with a
spherically symmetric metric. The dilaton field in the resulting model is a
relic of the angular variables hidden in this procedure. Also, it belongs to
the class of models first discussed by Jackiw and Teitelboim \cite{jac84},
which lend themselves to a gauge theoretical formulation \cite{jac84,jac96}
of the problem.

The formation of black holes in bidimensional dilaton gravity coupled to
scalar field was considered recently in different contexts. For instance,
the scalar matter field was already considered in the form of sine-Gordon
\cite{sso95} and quartic potentials \cite{jss95}, and these investigations
have motived us to introduce new systems for the scalar matter.

In this paper we investigate the formation of black hole in two different
systems. The first system considers a sixth-order potential for the matter
field that couples to the bidimensional dilaton gravity. This is a new
possibility, and although we yet have polynomial potential, the nature of the
solutions are different from the one that appear with the quartic potential
already investigated, since in the sixth-order potential the kink connects
the symmetric vacuum to an asymmetric one. The second system we shall deal
with is a system of two coupled fields recently introduced \cite{bds95}.
This system presents very interesting properties \cite{brs96,bba97}, and here
we show that the second field adds further effects.

This work is organized as follows. In the next Sec.~{\ref{sec:2}} we present
the CGHS model with a single field system, given by a sixth-order potential,
in Sec.~{\ref{sec:2.1}}.  The two field system as the source of matter is
worked out in Sec.~{\ref{sec:2.2}} and again we have black hole solutions, now
controlled by an enlarged number of parameters. Discussions and final remarks
are introduced in Sec~{\ref{sec:3}}.

\section{Models of Matter Fields}
\label{sec:2}

In this section we deal with two different examples of coulping dilaton
gravity to matter fields. The first example represents a single field system
that contains up to sixth power self-interaction terms. The second example is
different since it contains two coupled scalar fields, and is defined by a
potential that contains up to the quartic power in the fields, and presents
an enlarged set of parameters.

\subsection{System with a single field}
\label{sec:2.1}

Here we consider the model described by the action
\begin{eqnarray}
S&=&\frac{1}{2\pi}\int {d^2x}\sqrt{- \bar g}\Biggl\{\, 
e^{-2\phi}\Bigl[\bar R+4(\bar\nabla\phi)^2+4\lambda^2\Bigr]
\nonumber\\
& &-\frac{1}{2}(\bar\nabla f)^2+2\mu^2 f^2(f^2-a^2)^2 e^{-2\phi}\,
\Biggr\}~,
\end{eqnarray}
where $\bar g$, $\phi$ and $f$ are the metric, dilaton and matter fields
respectively. $\bar R$ is the scalar curvature and $\lambda^2$ is a
cosmological constant. This action is the usual action, except that the last
term contains a specific sixth-order potential for the self-interacting
matter field. 

We can be write Eq.~$(1)$ in a different form,
using a rescaled metric tensor $g_{\mu\nu}$ in such a way that
\be
{\bar g}_{\mu\nu}=e^{2\phi} g_{\mu\nu}~.
\ee
In this case, the action given by Eq.~$(1)$ turns into
\begin{eqnarray}
S &=& \frac{1}{2\pi}\int {d^2x}\sqrt{- g}\Bigl[e^{-2\phi} R+
4\lambda^2\nonumber\\& & -\frac{1}{2}(\nabla f)^2+
2\mu^2 f^2(f^2-a^2)^2\,\Bigr]~.
\end{eqnarray}
Note that this conformal reparametrization of the field eliminates the
kinetic term for the dilaton that appears in the original action.

The equations of motion that follow from Eq.~$(3)$ are
\be
\nabla^2(e^{-2\phi}) - 4\lambda^2 - 2\mu^2 f^2(f^2-a^2)^2=0~,
\ee
and 
\be
R=0~.
\ee
Equation~$(5)$ implies that $g_{\mu\nu}=\eta_{\mu\nu}$ and using the fact
that in two dimensions we can always put the metric in the conformal gauge
\be
{\bar g}_{\mu\nu}=e^{2\rho} \eta_{\mu\nu}~,
\ee
we conclude that $\rho=\phi$.

In order to investigate the solutions of this model it is useful
to introduce light-cone coordinates $x^{\pm}=t\pm x$. In these coordinates
the line element constructed with the metric tensor $g_{\mu\nu}$ is given by
\be
ds^2=-dx^+dx^- ~.
\ee
In terms of these coordinates, the action given by Eq.~$(3)$ can be cast
to the form 
\begin{eqnarray}
S&=&\frac{1}{\pi}\int {d^2x}\Biggl[\,\left( 2e^{-2\phi}\partial_+
\partial_-\rho+\lambda^2\right)\nonumber\\
& &\qquad-\frac{1}{2}\partial_+f\partial_-f-\frac{1}{2}
\mu^2 f^2(f^2-a^2)^2 \Biggr].
\end{eqnarray}
The field equations of motion are given by
\begin{eqnarray}
\partial_+^2\left(e^{-2\phi}\right)+\frac{1}{2}\left(\partial_+
f\right)^2&=&0,\\\partial_-^2\left(e^{-2\phi}\right)+
\frac{1}{2}\left(\partial_-f\right)^2&=&0,\\\partial_+\partial_
-\left(e^{-2\phi}\right)+\lambda^2-
\frac{1}{2}\mu^2f^2(f^2-a^2)^2&=&0,\\
\partial_+\partial_-f+\mu^2f(f^2-a^2)(3f^2-a^2)&=&0,
\end{eqnarray}
and these are the equations we have to deal with.These results follow very
naturally in the above procedure, and this should be contrasted to the
procedure used in Refs.~{\cite{sso95,jss95}} where one has to appropriately
choose the gauge to get to such r
esult. Here we see that the equation of motion for the matter field presents
solutions that can be cast to the form, working with standard coordinates
$(x,t)$
\be
f^2(x,t)=\frac{1}{2}a^2\{1+\tanh[\alpha((x-\bar{x})+v(t-\bar{t}))]\},
\ee
where $\alpha$ is a constant and $(\bar{x},\bar{t})$ represents the center of
the kink. In the light-cone coordinates we can write the above solutions as
\be
f^2(x^+,x^-)=\frac{1}{2}a^2\{1+\tanh[\alpha_+(x^+ -\bar{x}^+)-\alpha_-(x^-
-\bar{x}^-)]\},
\ee
where $\alpha_{\pm}=\frac{\alpha}{2}\left(v \pm 1\right)$.

In order to study black hole formation we use this $f$ into
the other equations of motion for $\phi$. Here we can write, for instance,
\be
\partial_+\partial_-\left(e^{-2\phi}\right)=\frac{\mu^2a^6}{16}\Bigl[1+\tanh^3\triangle-\tanh^2\triangle-\tanh\triangle \Bigr]-\lambda^2,
\ee
where we have set $\triangle=\delta-\bar{\delta}$, with $\delta=\alpha_+x^+
-\alpha_-x^-$ and $\bar{\delta}=\alpha_+\bar{x}^+ -\alpha_-\bar{x}^-$.
This equation can be integrated to give
\be
e^{-2\phi}=C_1+b(x^+)+d(x^-)-\lambda^2x^+x^- + \frac{\mu^2a^6}{16\alpha_+
\alpha_-}\Bigl[\frac{1}{2}\tanh\triangle-\ln\cosh\triangle\Bigr].
\ee
The functions $b(x^+)$ and $d(x^-)$ are determined by the constraint
equations, the two first equations of motion $(8)$ and $(9)$. Here we get
\begin{eqnarray}
b(x^+)&=&b\,x^+ +C_2,\\
d(x^-)&=&d\,x^- +C_3.
\end{eqnarray}
Therefore, the dilaton field can be determined up to constants in the form
\be
e^{-2\phi}=C+b\,x^+ +d\,x^- -\lambda^2x^+x^- + \frac{\mu^2a^6}{16\alpha_+
\alpha_-}\Bigl[\frac{1}{2}\tanh\triangle-\ln\cosh\triangle\Bigr],
\ee
where $b$, $d$, and $C$ are constants, and in the following we choose
$b=d=0$, for simplicity.

Let us now investigate the geometric nature of this solution generated by
original system. Toward this goal, let us divide spacetime into the three
regions: $\triangle=\delta-\bar{\delta}<<1$, $\triangle\approx 0$, and
$\triangle >>1$. In the first region the dilaton field becomes
\be
e^{-2\phi}\approx C-\lambda^2\left(x^+ +\frac{\mu^2a^6}{16\lambda^2\alpha_+}
\right)\left(x^- -\frac{\mu^2a^6}{16\lambda^2\alpha_-}\right)~,
\ee
where the constant $C$ is taken as
\be
C=\frac{\mu^2a^6}{16\alpha_+\alpha_-}\left(\bar{\delta}+
 \frac{\mu^2a^6}{16\lambda^2}-\ln2-\frac{1}{2}\right).
\ee

In the region where $\triangle=\delta-\bar{\delta}\approx0$ we get
\be
e^{-2\phi}\approx \frac{\mu^2a^6}{16\alpha_+\alpha_-}\left(
\bar{\delta}+\frac{\mu^2a^6}{16\lambda^2}-\ln2-
\frac{1}{2}\right)-\lambda^2x^+x^-.
\ee
With regard to the initial metric ${\bar g}_{\mu\nu}$, these solutions
represent, where  this metric is defined [1], the linear dilaton vacuum.

In the third region we have $\triangle=\delta-\bar{\delta}>>1$, and now
we obtain 
\be
e^{-2\phi}\approx \frac{\mu^2a^6}{16\alpha_+\alpha_-}(2\bar{\delta})
-\lambda^2\left(x^+ -\frac{\mu^2a^6}{16\lambda^2\alpha_+}\right)
\left(x^- +\frac{\mu^2a^6}{16\lambda^2\alpha_-}\right),
\ee
which represents, with respect to the original background,the geometry
of a black hole with mass
\be
\frac{\lambda\mu^2a^6}{8\alpha_+\alpha_-}\bar{\delta},
\ee
after shifting $x^+$ by $\mu^2a^6/16\lambda^2\alpha_+$
and $x^-$ by $-\mu^2a^6/16\lambda^2\alpha_-$.

\subsection{System with two fields}
\label{sec:2.2}

Let us now consider another system, defined by
\begin{eqnarray}
S&=&\frac{1}{2\pi}\int {d^2x}\sqrt{-g}\Biggl\{\, 
\Bigl[e^{-2\phi} R++4\lambda^2\Bigr]
\nonumber\\ & &-\frac{1}{2}(\nabla f)^2-\frac{1}{2}(\nabla g)^2+
U(f,g),\Biggr\}~.
\end{eqnarray}
This action can be written in terms of ${\bar g}_{\mu\nu}$ by doing the
inverse transformation that correspondes to Eq.~$(2)$.

Using light-cone coordinates the above action can be cast to the form 
\begin{eqnarray}
S&=&\frac{1}{\pi}\int {d^2x}\Biggl[\, \left( 2e^{-2\phi}\partial_+
\partial_-\rho+\lambda^2\right)
\nonumber\\ & &+\frac{1}{2}\partial_+f\partial_-f+
\frac{1}{2}\partial_+g\partial_-g -\frac{1}{4}U(f,g)\,\Biggr].
\end{eqnarray}
This new system is defined via the potential
\be
U(f,g)= \frac{1}{2}\mu^2(f^2-a^2)^2+\mu\nu(f^2-a^2)g^2+ \frac{1}{2}\nu^2g^4+
2\nu^2f^2g^2~.
\ee
This potential identifies a system of two real scalar fields
that was recently investigated in \cite{bds95}, where it was shown
to present interesting static field configurations. As we can see
from the above potential, we are now dealing with a richer system
and we want to explore how this enlarged system change the simpler
picture of black hole formation in systems of a single field.

We follow as in the former system.In this case we get
\begin{eqnarray}
\partial^2_+(e^{-2\phi})+\frac{1}{2}(\partial_+f)^2+\frac{1}{2}
(\partial_+g)^2&=&0~,\\
\partial^2_-(e^{-2\phi})+\frac{1}{2}(\partial_-f)^2+\frac{1}{2}
(\partial_-g)^2&=&0~,\\
\partial_+\partial_-(e^{-2\phi})+\lambda^2-\frac{1}{4}U(f,g)&=&0~,\\
\partial_+\partial_-f+\frac{1}{2}\mu^2(f^2-a^2)f+
\nu(\nu+\frac{1}{2}\mu\,fg^2&=&0~,\\
\partial_+\partial_-g+\frac{1}{2}\mu\nu(f^2-a^2)g+\frac{1}{2}\nu^2g^3+
\nu^2f^2g&=&0~.
\end{eqnarray}
The above Eqs.~$(31)$ and $(32)$ correspond to the matter field equations
of motion. They can be solved to give two different pair of
solutions \cite{bds95}:
\be
f=-a\tanh\{\mu a[\alpha_+(x^+-{\bar x}^-)-\alpha_-(x^--{\bar x}^-)]\},
\ee
and $g=0$, and also
\be
f=-a\tanh\{2\nu a[\alpha_+(x^+-\bar{x}^+)-\alpha_-(x^--\bar{x}^-)]\}~,
\ee
and
\be
g=\pm a\left(\frac{\mu}{\nu}-2\right)^{1/2}{\rm sech}\{2\nu a[\alpha_+
(x^+-\bar{x}^+)-\alpha_-(x^- -\bar{x}^-)]\}~,
\ee
valid for $\mu/\nu>2$. Here we note that the limit $\nu\to\mu/2$ transforms the
second pair of solutions into the first one, that presents $g=0$. This is
interesting since the investigation of the second pair of solutions allows
getting results valid for the first pair, with presents $g=0$ and so leads
to the case of just one field. See below for further details.

For the second pair of solutions we can cast the dilaton field in the form
\be
e^{-2\phi}=C-\lambda^2x^+x^- -\nonumber\\A\tanh^2[2\nu a(\delta-
\bar{\delta})]- B\ln\cosh[2\nu a(\delta-\bar{\delta})]~.
\ee
Like in the former case, in the above expression we have also chosen $b=d=0$
in $b(x^+)=b x^+ + C_2$ and $d(x^+)=d x^+ +C_3$, which follow from $(28)$ and
$(29)$. The expressions for $A$ and $B$ are given by
\be
A=\frac{a^2}{12\nu^2\alpha_+\alpha_-}\Biggl[\frac{\mu^2}{4}+
\frac{1}{4}(\mu-2\nu)(7\mu+10\nu)\Biggr]~,
\ee
and
\be
B=\frac{a^2}{12\nu^2\alpha_+\alpha_-}\Biggl[\mu^2+(\mu-2\nu)(\mu-
4\nu)\Biggr]~.
\ee

Let us now investigate the geometric nature of the black hole generated
in this system. Here we can write the dilaton field ,for $\Delta>>1$, as
\be
e^{-2\phi}\approx 4aB\nu{\bar\delta}
-\lambda^2 \left(x^+ -2\frac{\nu}{\lambda^2}Aa\alpha_-\right)
\left(x^- +2\frac{\nu}{\lambda^2}Aa\alpha_+\right)~,
\ee
which represents the geometry of a black hole with regard to
the original background spacetime.The mass of the black hole is given by
\be
4aB\lambda\nu\left({\bar\delta}+
\frac{\nu}{\lambda^2}Aa\alpha_+\alpha_-\right)~.
\ee
after shifting $x^+$ by $2(\nu/\lambda^2)aB\alpha_-)$ and $x^-$ by
$-2(\nu/\lambda^2)aB\alpha_+$. 

For $\Delta={\bar\delta}-\delta<<1$,
\be
e^{-2\phi}\approx
-\lambda^2 \left(x^+ +2\frac{\nu}{\lambda^2}Aa\alpha_-\right)
\left(x^- -2\frac{\nu}{\lambda^2}Aa\alpha_+\right)~,
\ee
where the constant $C$ was taken as
\be
C=A+2aB\nu{\bar\delta}-B\ln 2 -\frac{4\nu^2a^2B^2\alpha_+\alpha_-}{\lambda^2}~.
\ee
In the region of $\Delta\approx0$ we have
\be
e^{-2\phi}\approx
A-B\ln 2+2aB\nu{\bar\delta}-\frac{4a^2B^2\nu^2\alpha_+\alpha_-}{\lambda^2}
-\lambda^2x^+x^-~,
\ee
As in the first system, these last two solutions give us the linear
dilaton vacuum in the region where the original background is defined.

We recall that the limit $\nu\to\mu/2$ changes the second pair of
solutions that we have been considering to the first one, simpler, that
presents $g=0$, as introduced above. For this reason, we can investigate this
first pair of solutions by just setting $\nu\to\mu/2$ in the above results.

This procedure is interesting since it leads to the case with just one
field, more precisely to the case where the matter system is described by the
$\phi^4$ model, but this was already investigated in Ref.~{\cite{jss95}}.
Despite slightly different notations, it is not hard to see that the limit
$\nu\to\mu/2$ correctly change the above results into the results given in
\cite{jss95} for the case of a single field, in the $\phi^4$ model for the
matter contents.

\section{Comments and Conclusions}
\label{sec:3}

In this work we have investigated bidimensional dilaton gravity coupled
to two different matter field systems. The first system
is a single field system that contains self-interactions
up to the sixth power. It is of the same kind of the sine-Gordon \cite{sso95}
and $\lambda\phi^4$ \cite{jss95} models already investigated. The results of
these papers, together with the present work show that the black hole
that appears is qualitatively the same, but with different masses that
depend on the parameters associated with the various solutions. The second
system is a system of two fields, and contains up to the fourth power in
the fields. This system is richer since it is defined in a enlarged space
of parameters, which contains the space of parameters of the $\lambda\phi^4$
system as a particular case. The soliton solutions that appear in this case
also contributes to the generation of black holes, but these black holes are
quantitatively different from black holes that appear in systems of a single
field, the difference being controlled by the enlarged set of parameters
that defines the two field matter system.

Evidently, one can have more examples, for instance investigating the case
where the scalar matter is described by the two field system that contains
up to the sixth power in the fields, as also investigated in \cite{bds95}.
We think that it is interesting to investigate this kind of mechanism for
different matter field potentials, in order to understand mechanisms of
formation of black holes within the context of dilaton gravity. Furthermore,
there is also the newer context of the Einstein-Maxwell-dilaton-axion
system \cite{gke94} with general dilaton coupling. And this naturally leads
to another interesting issue, that concerns investigating the coupling of a
dilaton-axion gravitational field with two coupled scalar fields, in the form
introduced in \cite{bds95,brs96}. The two cases worked out above lead us to
different scenario concerning formation of black holes, and so it turns out
to be interesting to analyze quantum or at least semiclassical versions
\cite{jac96} since these black holes may present new termodinamics properties,
namely the Hawking-Bekenstein radiation. Work on this and in other related 
issues is now in progress.

\vskip 1cm


D.B. would like to thank Roman Jackiw for comments and for reading the
manuscript.

\end{document}